\documentclass[final,5p,times,twocolumn]{elsarticle}

\usepackage{graphicx}

\usepackage{amssymb}

\journal{Physics Letters A}

\begin{document}

\begin{frontmatter}



\title{Asymmetrical solutions and role of thermal fluctuations in dc current driven extended Josephson junction. }

\author{A. N. Artemov}
\ead{artemov@fti.dn.ua}
\address{Donetsk Physical and Technology Institute, Donetsk 83114, Ukraine}

\begin{abstract}
Extended Josephson junction driven by dc bias current is studied numerically.
Two types of solutions, symmetrical and asymmetrical, are found. The current-voltage characteristic (IVC) is calculated. The symmetrical solutions form main histeretic IVC and asymmetrical one create an additional branch. Depending on the bias current value periodic, quasiperiodic and chaotic modes of the junction motion was observed. Dynamics of the junction affected by thermal fluctuations was analyzed. Stability of different states of the junction is discussed.

\end{abstract}


\end{frontmatter}
\section{Introduction}
\label{s1}

Extended Josephson junction (EJJ) driven by dc current is, from the theoretical standpoint, a nonlinear dynamical system. Mathematical model to describe the EJJ is the damped sine-Gordon equation \cite{barone}. Depending on the current value the EJJ can be in a statical or a dynamical state.
In the wide range of the damped parameter the junction IVC is hysteretic. In this case at some values of bias current the EJJ can be in one of two different states. In the absence of external perturbations each of these states is stable. But thermal fluctuations can switch the junction from one state to another \cite{han}.

Stability of the junction states under thermal fluctuations is special interest in last years. Authors of the work \cite{gas} have investigated the rate of thermally induced escape from the zero-voltage state in long Josephson junctions (LJJ). Return to zero-voltage state current statistics under thermal fluctuation was analyzed in \cite{leo}. Lifetime of superconducting states of long and short JJs was investigated in works \cite{fed1, fed2, aug}.

The aim of this work is detailed study of the dc current driven EJJ in the in-line geometry without external magnetic field. Symmetrical and asymmetrical solutions of corresponding sine-Gordon equation were found numerically. The hysteretic IVC of the junction formed by solutions found was calculated. The asymmetrical solutions form the additional branch of the IVC. Depending on the dc bias current value the EJJ can demonstrate different types of motion. Periodic, quasiperiodic and chaotic behavior of the junction were revealed. The effect of thermal fluctuations on the stability of symmetrical and asymmetrical states was investigated.

\section{Mathematical model}
\label{s2}
To model the EJJ the two-dimensional (2D) damped sine-Gordon model was used \cite{barone}
\begin{equation}\label{e2.1}
    \frac{\hbar C}{2e}\frac{\partial^2 \varphi}{\partial t^2}
    +\frac{\hbar g}{2e}\frac{\partial \varphi}{\partial t}+j_c\sin \varphi=
    \frac{c^2\hbar}{8\pi e d}\left(\frac{\partial^2 \varphi}{\partial x^2}+\frac{\partial^2 \varphi}{\partial y^2}\right) + \frac{2e}{\hbar}F.
\end{equation}
Here $\varphi$ is the phase difference of order parameters of the superconductors forming the junctions, $j_c$ is the critical current density of the Josephson junction, $\hbar$ is the Planck constant, $e$ is the value of the electron charge, $c$ is the light speed, $C$ and {g} are the capacity and the conductivity of the junction per unit aria, $d$ is the thickness of the junction. The random force $F$ conjugated with the phase difference simulates effect of thermal fluctuations on the junction. It is the Gaussian $\delta$-correlated random variable
\begin{equation}\label{e2.2}
    \langle F(\textbf{r},t)F(\textbf{r}',t')\rangle=\sigma^2\delta(\textbf{r}-\textbf{r}')\delta(t-t')
\end{equation}
with the dispersion
\begin{equation}\label{e2.3}
    \sigma^2=2T\frac{\hbar^2 g}{(2e)^2}.
\end{equation}

In this work for the sake of simplicity the model junction in the form of an infinite $L$-wide strip is considered. The strip is directed along the $y$ coordinate. The edges of the strip are defined as $x=\pm L/2$. The boundary conditions in the case of a dc current driven junction in the in-line geometry are given by
\begin{equation}\label{e2.4}
    \frac{\partial \varphi}{\partial x}\left(x=\pm \frac{L}{2},y,t\right)=\pm\frac{4\pi^2d}{\Phi_0c}I,
\end{equation}
where $\Phi_0$ is the magnetic flux quantum and $I$ is the value of the dc bias current per unit length of the junction. In the $y$ direction the periodic boundary conditions with the period $L$ are supposed.

To obtain dimensionless expressions $\tau=(\hbar C/2ej_c)^{1/2}$ was used as the time unit, $\lambda_J=\sqrt{\Phi_0c/(8\pi^2dj_c)}$ was used as the length unit, voltage is measured in $2e/\hbar\tau$ and current was normalized by $\lambda_JLj_c$. In order to solve the equation with the random force numerically it was rewritten in the form of two differential equations of the first order in time
\begin{eqnarray}\label{e2.5}
   \varphi_t&=&u ,\\ \label{e2.6}
   u_t&+&\gamma u+\sin \varphi=\varphi_{xx}+\varphi_{yy} + \tilde{\sigma}f_0
\end{eqnarray}
with the boundary conditions
\begin{eqnarray}\label{e2.7}
\!\!  \varphi_x(\pm L/2,y,t)=\pm I/2, &&\varphi(x,0,t)=\varphi(x,L,t) \\
\!\!  u(\pm L/2,y,t)=\varphi_t(\pm L/2,y,t), \;&&u(x,0,t) = u(x,L,t). \label{e2.8}
\end{eqnarray}
Here $u(\textbf{r},t)$ is the dimensionless local voltage on the junction, $\gamma$ is the dimensionless damping parameter, $f_0$ is the Gaussian random variable with the unit dispersion and the dimensionless dispersion is given by the expression
\begin{equation}\label{e2.9}
    \tilde{\sigma}^2=4\gamma\frac{8\pi T}{H_J^2(T)\lambda_J^2(T)d(T)}=
    4\gamma \frac{\tau}{\sqrt{1-\tau^2}}\sqrt{Gi},
\end{equation}
where $H_J=\Phi_0/\pi\lambda_J d$ is the overheating field of the EJJ, $\tau=T/T_c$, $T_c$ is the superconducting transition temperature of the metal forming the junction. To extract the temperature factor in eq.(\ref{e2.9}) the model temperature dependence of the London length $\lambda_L(T)=\lambda_L(0)/\sqrt{1-\tau^2}$ was supposed. The Ginzburg-Levanyuk parameter \cite{lev,gin} can be written in the form
\begin{equation}\label{e2.10}
    Gi=\left(\frac{8\pi T_c}{H_J^2(0)\lambda_J^2(0)d(0)}\right)^2.
\end{equation}
It characterizes the dependence of the random force simulating thermal fluctuations on material parameters of a junction. To gain some insight of this parameter value we evaluated it for the lead junction. Taking into account that in this case $T_c=7.2$K and $\lambda_L(0)=37$nm we find $Gi\thickapprox 2 \cdot 10^{-9}$.

In the absence of fluctuations the model considered is identical to the LJJ because its solutions don't depend on the coordinate $y$. But the dispersion (\ref{e2.3}) due to its dimension corresponds to the 2D space and 1D time distributed random force. To consider a long $(L\gg\lambda_J$) and narrow ($w<\lambda_J$) junction one has to average  eq.(\ref{e2.1}) over the width. Corresponding dispersion depends on $w$. But in this case in the in-line geometry we have to use more complicated boundary conditions.

To solve eqs.(\ref{e2.7}-\ref{e2.8}) the explicit finite difference scheme of the first order in time and the second order in space was used. The time interval was accepted $\triangle t=0.01$ and the space interval was $\triangle x=0.1$.

\section{Main branches of IVC without fluctuations}
\label{s3}

In this work we intend to discuss the influence of thermal fluctuations on a behavior of the EJJ. Intensity of the fluctuations depends on temperature. In our model this fact is taken into account as the temperature dependence of the random force dispersion (\ref{e2.9}). In this situation the effective junction width $L$ depends on temperature because
it is measured in the unit $\lambda_J(T)$. The time independent width of the junction considered is $L_0=14.14\lambda_J(0)$. At temperature $T=0.5T_c$ which is accepted in this work the effective width is $L= 10$.

At first, we examine the EJJ behavior in the absence of fluctuations. In this case solutions of the EJJ don't depend on the $y$ coordinate and in fact the junction is long. To get overview about dynamical states of the contact we have calculated its IVC. We consider the IVC as a diagram in the $I-\langle u\rangle$ plane, where $\langle u\rangle$ is the dc voltage, each point of which depicts the attractor of the dynamical system. The diagram allows us to systemize the attractors with regard to a way in which it may be reached by the system.

To calculate IVC we start from the zero initial conditions at current value less then critical one, which is $I_c=4$. Then we increased it by small steps up to maximum value to obtain the ascending branch of the IVC. In a similar way starting  from the attractor at highest current value and decreasing current step by step up to reaching a zero voltage state, the descending branch was obtained. Solutions at the previously calculated nearby point are taken as the initial conditions for the next point. After each changing of the current the system during $t=500$ was relaxed to a new attractor. Then the dc voltage $\langle u\rangle$ on the junction was calculated as the averaging of the value of the local voltage $u(\textbf{r},t)$ over $L\times L$ space area and $t=1000$.
\begin{figure}[h]
  \includegraphics[width=8.5cm]{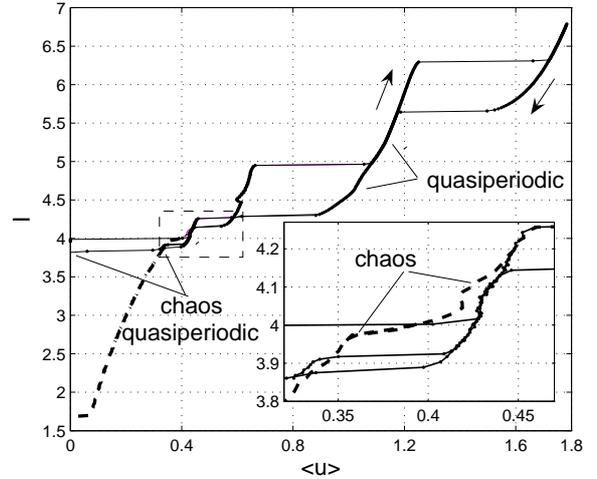}\\
  \caption{IVC of the EJJ. Arrows show the the direction of the current varying on the IVC branches. Dashed line plots the separated branch of the IVC. Insert shows the portion of the IVC confined by the dashed rectangle. The segments of the branches, where quasiperiodic and chaotic types of motion are observed, are pointed.}
  \label{f1}
\end{figure}

Resulted IVC is shown in fig.\ref{f1} by closed circles. As it follows from the way in which the IVC was obtained, all these attractors can be reached starting from zero initial conditions.
The IVC demonstrates hysteresis behavior. A typical peculiarity of such IVC is that the return into zero voltage state current is less than critical one. For EJJ discussed it is $I\approx 3.837$.

Solutions of eqs.(\ref{e2.5}-\ref{e2.6}) positioned on the main IVC possess the space symmetry specified by the zero initial conditions and the boundary conditions  (\ref{e2.7}-\ref{e2.8}). That is the phase difference and the voltage distributions are symmetrical with respect to the center of the junction ($x=0$) and the magnetic field distribution is antisymmetrical. Such solution are well known \cite{barone,fil}.

To classify the type of the dynamical system motion at the given bias current $I$ the phase portraits $u(\textbf{r},t)$ vs $\cos\varphi(\textbf{r},t)$ was plotted, were $\textbf{r}$ is the space coordinate positioned in the center of the junction.
In the most points of the IVC the motion of the system is periodic. The attractors of such type of the motion are closed lines. We will not discuss this type of motion in more details here.

To analyze more complicated motion we calculated the time series $v_i=\langle u(\textbf{r},t_i)\rangle$ at given current $I$. Each term of the series is the voltage averaged over $L\times L$ junction area calculated with the time interval $\tau=1$. Each such series contains 5000 terms.

\begin{figure}[h]
  \includegraphics[width=8.5cm]{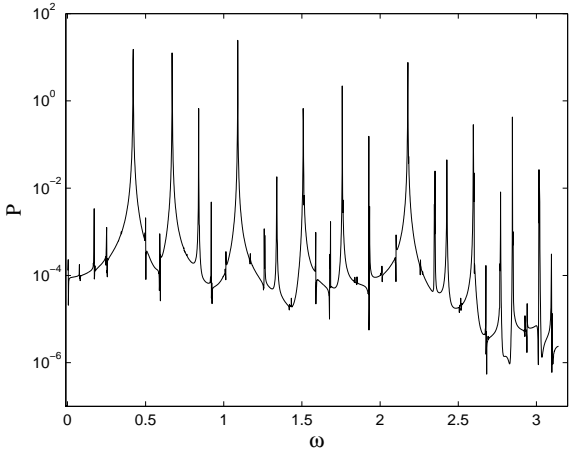}\\
  \caption{Power spectrum of the quasiperiodic motion at the bias current value $I=5.0$.}
  \label{f2}
\end{figure}

Complicated multi-frequency  motion of the system is observed in the current interval $I\in (4.62-5.26)$ on the descending branch of the IVC. The trajectory on the phase portrait visually fills up the whole region.
The power spectrum of the voltage for such motion is shown in fig.\ref{f2} and it demonstrates multitudinous spectrum lines. The evaluation of the correlation dimension gives $D_2=2$ what is an argument for a quasiperiodical motion. Such a motion consists of periodic motions with incommensurate frequencies.

\begin{figure}[h]
  \includegraphics[width=8.5cm]{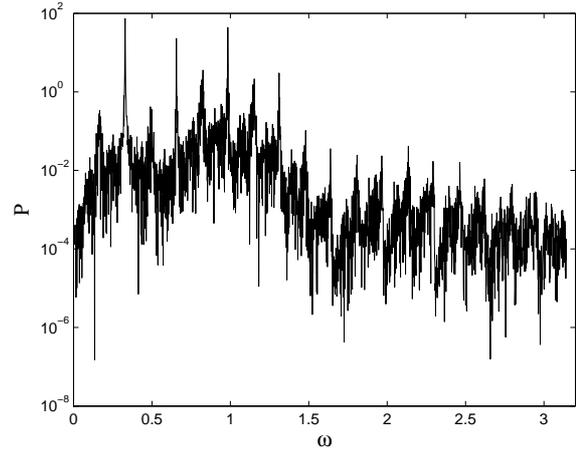}\\
  \caption{Power spectrum of the chaotic motion at the bias current value $I=3.87$.}
  \label{f3}
\end{figure}

Rather complicated picture is observed in the current interval from $I=3.881$ to the return current $I=3.837$. Here narrow current intervals where the system demonstrates a chaotic behavior are intermited by those with quasiperiodic motion. Thus, at currents $I=3.881-3.865, 3.854, 3.849, 3.844, 3.840$ the chaotic motion was found, whereas at
$I=3.864-3.855, 3.852, 3.847, 3.842$ the quasiperiodic one is observed.
The power spectrum of the chaotic motion at current $I=3.87$ is shown in fig.\ref{f3}. It is characterized by the wide band of a continuous spectrum what is typical for a chaos. The correlation dimension of the strange attractor is evaluated to be $D_2=1.73$.

All attractors placed on the main IVC are joined by the fact that they can be reached starting from symmetrical zero initial conditions. These conditions are similar to those which are take place in real experiments with EJJ driven by dc bias current. And this IVC contains all possible dynamical states of the junction which can be reached by means of changing bias current without additional external influences. But the dynamical system (\ref{e2.5}-\ref{e2.6}) has another attractors which don't contain symmetrical initial conditions in their domains of attraction. Some of them will be discussed in the next section.

\section{Separated branch of IVC}
\label{s3}

In previous section the main branches of IVC were considered. Here an additional separated set of attractors are discussed.
This branch is shown in fig.\ref{f1} by dashed line. It is placed in the low bias current ($I\in (1.7-4.16)$) region of the IVC. Nonzero voltage states can be stable at current much less than critical on the branch.
The system can be moved reversibly along the whole branch by means of bias current varying excluding the highest and the lowest current points where the system switches to the main branch.

Solutions of eqs.(\ref{e2.5}-\ref{e2.6}) on this branch are asymmetrical in contrast to that on the main IVC. Domains of attraction of these attractors don't contain symmetrical (zero) initial conditions. Asymmetrical initial conditions must be used to obtain such solutions. Thus, the initial phase difference distribution $\varphi(x)=0.2x^2+0.6x+0.4$ and zero voltage distribution fall into the domain of attraction of the attractors on the separated branch corresponding to the current interval $I\in (3.7-4.0)$.

\begin{figure}[h]
  \includegraphics[width=8.5cm]{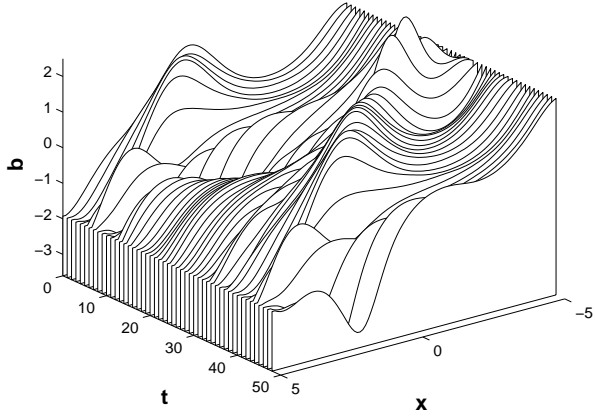}\\
  \caption{Magnetic field space-time evolution at $I=3.85$ on the separated branch of the IVC.}
  \label{f4}
\end{figure}

The solutions in the upper segment of the branch ($I>4$) correspond to states in which two fluxons enter in the junction from opposite edges in different moments of time and annihilate inside of the junction. On the lower segment one only fluxon is situated in the junction. It enters from left edge, moves to right and exits (annihilates) from right edge. Then the fluxon of opposite sign enters through right edge and moves to left. Space-time evolution of such type of solution is shown in fig.\ref{f4}.

 On the most part of the branch a periodic motion is observed. In this case the period of motion consists of two stages of entering and annihilation of fluxons.

\begin{figure}[h]
  \includegraphics[width=8.5cm]{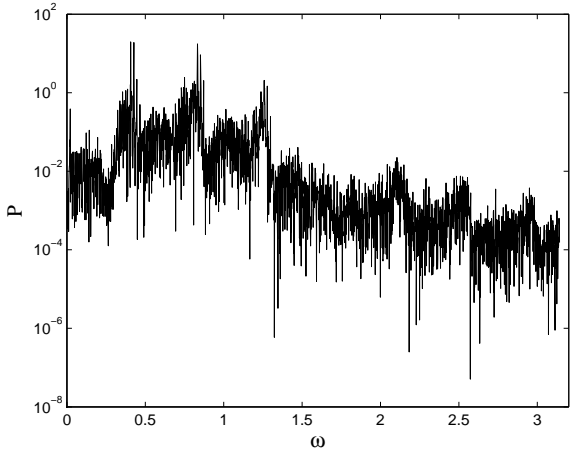}\\
  \caption{Power spectrum of the chaotic motion at the bias current value $I=4.02$ on the separated branch.}
  \label{f6}
\end{figure}

 The system demonstrates also a chaotic behavior in the current interval $I\in (3.98-4.10)$. The power spectrum corresponding to the bias current value $I=4.02$ is shown in fig.\ref{f6}. The correlation dimension of the attractor at such current is $D_2=1.86$.

Existence of the separated branch of attractors makes it possible for the EJJ to demonstrate different types of motion at the same value of current. We focus your attention on the fact that in the current interval $I\in (3.98-4.10)$ the system demonstrates periodic type of motion on the main branch of the IVC and chaotic one  on the separated that.

\section{Influence of thermal fluctuations on the JJ behavior}
\label{s4}

Usually the influence of thermal fluctuations on the dynamical behavior of a system depends on the ratio between typical sizes of an attractor and fluctuations. If the size of fluctuations is small in comparison with the attractor one the dynamical motion can be separated from noise and analyzed.  In the opposite case fluctuations destroy the attractor and noise will predominate.

To investigate the influence of thermal fluctuations on the dynamical properties of the EJJ we solved eqs.(\ref{e2.5}-\ref{e2.6}) numerically taking into account the random force. The intensity of model fluctuations was defined by the parameter $Gi=10^{-7}$ and the reduced temperature $\tau=0.5$.

Results of the calculations can be summarized as the following. Characteristic sizes of the fluctuations under conditions given are noticeably less than attractor ones. So, the system moves in the vicinity of the attractor corresponding to given bias current. Dc voltage on the IVC branches remains practically the same as without fluctuations.

But presence of the separated set of attractors made the system behavior to be very sensitive to the fluctuations influence in the bias current interval $I\in (3.84-4.10)$ where two solutions of eqs.(\ref{e2.5}-\ref{e2.6}), symmetrical and asymmetrical, can be realized at each value of current. It is found that under given conditions the lifetime of symmetrical solutions (on the main IVC) are much less than that on the separated branch. The thermal fluctuations are large enough to push out the system during finite time ($t\sim10^3$) from a position on the main branch into the domain of attraction of the attractor on the separated one. At higher currents ($I>4.1$) the stability of the system states on the main IVC rises steeply. Switches of the states was not observed during $t=2\cdot10^5$.

\begin{figure}[h]
  \includegraphics[width=8.5cm]{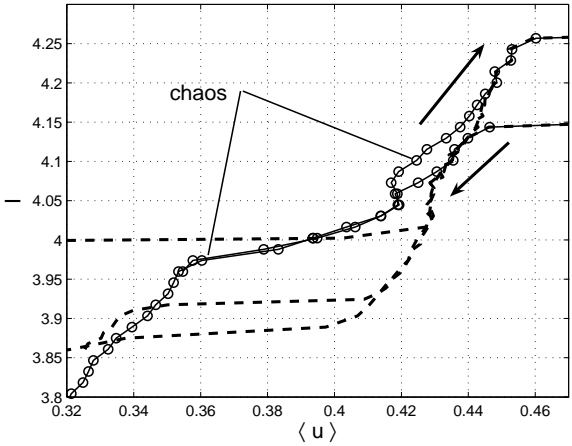}\\
  \caption{Dependence of $I$ vs $\langle u \rangle$ for EJJ taking into account thermal fluctuations. Open circles mark ascending and descending branch of the actual IVC. Dashed lines plot the branches of fluctuationless main IVC. }
  \label{f5}
\end{figure}

One of the possible variants of actual IVC is shown in fig.\ref{f5} by lines marked by open circles. It was calculated in the same way as before. The picture represents the reduced portion of the IVC on which difference between actual and fluctuationless IVC is visible. At more high values of the bias current the IVC practically doesn't depend on fluctuations.

The process of the system switching from one metastable dynamical state to another is probabilistic.
It means that actual IVC will depend on the regime of calculations or measuring. During calculations under regime chosen ($t=500+1000$ for each point) we don't observe revers system hops, from the separated into the main branch. The more detailed examination of the states stability (calculation time $t\sim 5\cdot10^5$) doesn't establish revers hops too. It means that the lifetime of the system on the separated branch is $\tau_s\gg 10^5$ whereas that at main one is $\tau_m\sim 10^3-10^4$.
Switches between metastable states on the ascending and the descending branches of the main IVC also was not observed.

\section{Conclusion}
\label{s5}

Nonlinear systems at given value of driving parameter can have several attractors separated by separatrix in a phase space. To locate the system on a partial attractor it is need to specify initial conditions within its domain of attraction which often is unknown.

Well known sample of such a system is the Josephson junction with hysteretic IVC discussed above. At some values of the bias current the system can occupy one of two attractors. A state of such a junction at given current depends on the way to obtain this state. All solutions on the main IVC possess the symmetry specified by the antisymmetrical boundary and the symmetrical initial conditions.

In this work the additional set of attractors is found. It termed as the separated branch of the IVC. Corresponding solutions are asymmetrical. To find them asymmetrical initial conditions, similar to one-fluxon solutions, must be used.

In the absence of external perturbation the solutions on both the main and the separated brunches are stable. Under fluctuations affect the situation is different. If at given bias current value the system allows more than one solution, they become unstable for switching between them. It is found that under conditions chosen ($Gi=10^{-7}$, $T=0.5T_c$) the lifetime of the system on the main branch at current $I<4.1$ is quite small $\tau_m\sim 10^3$ in comparison with that on the separated one $\tau_s\gg 5\cdot10^5$. That is, in these conditions the system state on the separated brunch is preferable. At current $I> 4.1$ switching between branches was not observed during time period $t=2\cdot10^5$.

Switching between the separated and the main branchs reveals itself in the change of values of dc voltage and return current. Moreover, as a result of such a switch a type of motion can be changed. Thus, in the bias current interval $I\in (3.99-4.10)$ the EJJ switches from periodic attractor on the main IVC to chaotic that on the separated brunch.

The IVC considered (fig.\ref{f5}) is determined by the characteristic time ($t=500+1000$) spent to calculate one point on the IVC. This time is rather large as a calculation one but it is much less then a characteristic time of a dc current experiment. The behavior of a real junction will depend on the fluctuations intensity (parameter $Gi$, temperature) and details of the experiment. It can vary from a steady dynamical state on the branch of the main IVC to a telegraph noise signal with two or even three levels of voltage.

\end{document}